\documentclass[aps,pra,showpacs,reprint]{revtex4-1}
\usepackage{amsmath}
\usepackage{amssymb}
\usepackage{graphicx}
\usepackage{bbm}
\usepackage[colorlinks=true,urlcolor=blue,menucolor=blue,citecolor=black,linkcolor=blue]{hyperref}
\usepackage{xcolor}

\newcommand{\ket}[1]{|{#1}\rangle}
\newcommand{\bra}[1]{\langle{#1}|}
\newcommand{\tr}{\mathrm{Tr}}

\begin{document}
\title{Optimal choice of state tomography quorum formed by projection operators}
\author{Violeta N. Ivanova-Rohling}
\affiliation{%
Department of Mathematical Foundations of Computer Sciences, Institute of Mathematics and Informatics, Bulgarian Academy of Sciences,
Akad.G.Bonchev, block 8,
1113 Sofia, Bulgaria}
\email{violeta@math.bas.bg}
\author{Niklas Rohling}
\affiliation{Center for Quantum Spintronics, Department of Physics, Norwegian University of Science and Technology, NO-7491 Trondheim, Norway}
\email{niklas.rohling@ntnu.no}
\begin{abstract}
 A minimal set of measurement operators for quantum state tomography has in the non-degenerate case ideally eigenbases which are mutually unbiased.
 This is different for the degenerate case.
 Here, we consider the situation where the measurement operators are projections on individual pure quantum states.
 This corresponds to maximal degeneracy.
 We present numerically optimized sets of projectors and find that they significantly outperform
 those which are taken from a set of mutually unbiased bases.
\end{abstract}

\maketitle

\section{Introduction}
Any physical system which is supposed to function as a building block of a quantum computer requires a procedure to determine its state
in order to demonstrate its functionality and if necessary to debug it.
The measurements and computations which allow estimating a quantum state is called {\it quantum state tomography}.
Therefore, it has been performed on
trapped ions \cite{Roosetal2004,Haeffneretal2005,Haljanetal2005,Homeetal2006,Haeffneretal2008},
photonic qubits \cite{OBrienetal2003},
superconducting qubits \cite{Liuetal2005,Steffenetal2006PRL,Steffenetal2006Science,Neeleyetal2008},
spin qubits in quantum dots \cite{Folettietal2009,Shulmanetal2012,Medfordetal2013,Watsonetal2018,Zajacetal2018},
$^{13}$C and N nuclear spins at a nitroge-vacancy defect in diamond \cite{Waldherretal2014}.

For an $n$-dimensional Hilbert space, the density matrix has $n^2-1$ parameters, which need to be estimated.
This can be achieved by projective measurements, i.e., for known states it is counted by repetitive measurements
how often the unknown state is projected onto them.
In the non-degenerate case, one observable can provide projections on $n$ eigenstates, from which $n-1$ are actually useful.
A minimal set of observables which provide knowledge about the complete density matrix is called \textit{quorum} \cite{ParkBand1971}.
It contains $n^2-1$ distinct states.
As a fixed finite number of measurements can only provide an estimate for the quantum state, a central question of state tomography is
how to choose the measurements such that these estimates are as precise as possible.
For non-degenerate measurement operators the ideal choice of the quorum corresponds
to mutually unbiased bases (MUBs) \cite{WoottersFields89},
i.e. the eigenstates of the measurement operators form such bases.
"Mutually unbiased" means that a measurement outcome in one bases, or for one of the operators,
does not reveal any information about the other measurements.
Note that in general, state tomography is realized by positive-operator valued measures (POVMs), see e.g.
\cite{Renesetal2004,Rehaceketal2004}.
However, in this paper, we will restrict the discussion to projective measurements.
Further note that if loss in the measurement process has to be taken into account,
which effectively refers to $\tr\rho<1$, the optimal choice of the measurement set is different from MUBs
\cite{Miranowiczetal2014,Miranowiczetal2015,Bartkiewiczetal2016}.
In that situation $n^2$ parameters have to be determined.

If some of the restrictions to a minimal predefined measurement set are dropped,
more possibilities for finding an optimal tomography scheme exist.
This holds if the tomography scheme can include more than $n^2-1$ quantum states \cite{RoyScott2007,deBurghetal2008} or
the states can be changed after some of the measurements are done so that the quantum state is not completely unknown and
the measurements can be adapted \cite{HuszarHoulsby2012}.

In this article, we will keep the restrictions to exactly $n^2-1$ states and the decision about them before the start of the measurement,
which is relevant for experiments where it is difficult to alter the measurement setting.
However, we will not consider the non-degenerate case but the situation where each measurement operator is a projector on one pure quantum state.
Then, a quorum consists of $n^2-1$ such projectors.
Previously, projectors on basis states from a set of MUBs have been suggested \cite{RohlingBurkard2013} for such a situation.
Their performance is clearly better than a quorum based on non-entangled states only \cite{James_et_al2001}.
However, by applying a numerical search, we show in this article that there are quorums which perform even better than MUBs.
This is possible due to the fact that the involved $n^2-1$ states can be freely chosen while in MUBs the states within
one of the bases have a fixed, non-ideal relation to each other.

The article is organized as follows:
In Sec.~\ref{physsys}, we discuss physical implementation for which our considerations are relevant, namely
measurements based on spin-to-charge conversion in a double quantum dot, see Sec.~\ref{spin-to-charge},
and measurements of photonic orbital angular momentum with only two photon counters, see Sec.~\ref{quantumoptics}.
Sec.~\ref{searchspace} provides the formalization of the situation as an optimization problem, which is solved numerically in
Sec.~\ref{numerics} including a discussion of the results.
We compare the obtained results to the performance of MUBs in Sec.~\ref{compare}
and present conclusions and an outlook on more general situations in Sec.~\ref{conclusionsandoutlook}.

\section{Physical systems with projective measurements on individual states}
\label{physsys}
The work which we present here is relevant for quantum systems which are measured by projections on individual quantum states,
i.e., the ,measurement is a projection on a one-dimensional subspace of the Hilbert space.
In the following we discuss two relevant implementations.

\subsection{Spin-to-charge conversion}
\label{spin-to-charge}
For two spin qubits stored in a double quantum dot, i.e., one implementation of a four-dimensional Hilbert space,
electron spin resonance allows for single-qubit \cite{Koppensetal2006}
and the controlling the exchange interaction allows for universal two-qubit gates \cite{Pettaetal2005}.
Both types of operations have been demonstrated in the same double dot \cite{Brunneretal2011,Zajacetal2018}.
Thus these double dots could be a building block for a quantum computer.
Apart from the possibility to read out each spin qubit individually as done in \cite{Zajacetal2018},
spin-to-charge conversion \cite{Kane1998,Vandersypenetal2004} can be applied.
This means the electric potential at one of the dots is reduced so that both electrons will go to this dot
provided that the spin state of this two electron system ends up in a singlet state at the end of this sweep.
Measuring the charge state then projects either on one quantum state, if the charge of the energetically lower dot is measure to be two elementary charges,
or on the remaining three-dimensional subspace, if only one elementary charge is detected.
This is a realization in four dimensions of the situation we consider in this article.
Typically the state tomography is supposed to determine the density matrix of the two-qubit system before the sweep.
Note that the state which is connected to the singlet state after decreasing the electric potential in one of the dots
is not necessarily the spin singlet state, but depends actually on the speed of the transition
\cite{Folettietal2009,Shulmanetal2012,RibeiroBurkard2009,Ribeiroetal2010,RohlingBurkard2013}.

\subsection{In quantum optics}
\label{quantumoptics}
Nicolas et al,~\cite{Nicolasetal2015} described the realization of state tomography for a photonic qubit
where the quantum information is encoded in the orbital angular momentum of the light.
The measurements are projections onto certain qubit states which correspond to the x, y, and
z axes of the Bloch sphere.
The authors discuss potential extensions to higher dimensions.
In their proposed setups there are as many single-photon detectors included as the dimension of the quantum state.
This corresponds to the possibility to perform non-degenerate measurements.
However, if all but two detectors were removed,
one could in this then simplified setup obtain the measurements by projection on individual states.
One of the remaining detectors can detect a photon if the photonic qubit have been in a certain state while the
other one can detect it if it was in any state of the remaining $(n{-}1)$-dimensional subspace.
Actually the second detector is only needed to determine the ratio of detected to non detected photons,
while theoretically one detector would be sufficient.

\section{Defining the search space}
\label{searchspace}
We denote a state $\ket{\psi}$ in an $n$-dimensional quantum system by
\begin{equation}
\label{eq:staterepres}
\begin{split}
\ket{\psi} = & \cos\theta_1 \ket{1} + \sin\theta_1\cos\theta_2 e^{i\phi_2} \ket{2} + \ldots \\
             & + \sin\theta_1\sin\theta_2\ldots\cos\theta_{n-1}e^{i\phi_{n-1}}\ket{n-1}\\
             & + \sin\theta_1\sin\theta_2\ldots\sin\theta_{n-1}e^{i\phi_{n}}\ket{n},
\end{split}
\end{equation}
where $\{\ket{1},\ldots,\ket{n}\}$ is an orthonormal basis in the $n$-dimensional Hilbert space,
which describes our system.
Note that our state is fully given by the $2n-2$ real parameters $\{\theta_1,\theta_2,\ldots\theta_{n-1},\phi_2,\phi_3,\ldots,\phi_n\}$.
Here, we took into account the normalization of the state and that a different global phase does not yield a different physical state.

In general the state of the quantum system is described by a density matrix $\rho$,
which is a $n\times n$ Hermitian matrix with trace 1.
This means $n^2-1$ real parameters need to be determined by quantum state tomography.
As in the situation we consider here, each measurement is just a projection on a certain quantum state,
we need at least $n^2-1$ projection operators, which are linear independent of each other within the vector space
of $n\times n$ matrices.
If we denote the quantum states which form a quorum, i.e. minimal set for state tomography,
by $\ket{\psi_1},\ldots,\ket{\psi_{n^2{-}1}}$,
then $\rho$ can be determined by obtaining experimental estimates of
\begin{equation}
 A_i = \tr(\mathcal{P}_i\rho)=\tr(\ket{\psi_i}\bra{\psi_i}\rho), ~~~~i=1,\ldots,n^2-1,
\end{equation}
from repeated measurements of the projection $\mathcal{P}_i=\ket{\psi_i}\bra{\psi_i}$.

As we know already that $\tr\rho=1$, we can disregard the component of $\mathcal{P}_i$ which is 
proportional to $\mathbbm{1}$,
\begin{equation}
 \mathcal{Q}_i := \mathcal{P}_i-\mathbbm{1}/n.
\end{equation}

\subsection{Number of free parameters of the optimization problem}
As a quorum is formed by $n^2-1$ states and one state is determined by $2n-2$ real parameters,
the number of parameters, $N_{\rm param}$, for our optimization problem seems to be $2(n^2-1)(n-1)=2(n^3-n^2-n+1)$.
However, this number can be reduced by making use of the fact that any unitary transformation
of all of the states, leaves the resulting precision of the state tomography unchanged.
This is because we have assumed that all individual projective measurements can be done with
the same precision.
By using this fact, we can set fixed values for some of the states' parameters.
The effect is that in the optimization problem, which we formulate here,
the numbers of parameters is reduced by eliminating equivalent solutions, which are
those which are connected by a global unitary transformation.
This reduces the number of parameters by $n^2-1$, which is the dimension of
the special unitary group $SU(n)$, leading to
\begin{equation}
 N_{\rm param} = 2(n^3-n^2-n+1) - (n^2 -1) = 2n^3 - 3n^2 - 2n + 3,
\end{equation}
leaving the leading order to be cubic in $n$, see Table \ref{tbl:numfreeparams}.
\begin{table}
 \begin{tabular}{|l| l| l| l| l|l|l|}
 \hline
  $n$              & 3  & 4  &  5  &  6 & 7 & 8\\
  \hline
  $N_{\rm param}$  & 24 & 75 & 168 &315 &528& 819\\
  \hline
 \end{tabular}
 \caption{Number of free parameters, $N_{\rm param}$, for our optimization problem for dimension $n$.}
 \label{tbl:numfreeparams}
\end{table}

Practically, we can fix the parameters which we want to exclude by setting for the fist $n$ states, without loss of generality,
\begin{equation}
 \begin{split}
      \ket{\psi_1} = & \ket{1},\\
      \ket{\psi_2} = & \cos\theta_{21}\ket{1} + \sin\theta_{21}\ket{2},\\
      \ket{\psi_3} = & \cos\theta_{31}\ket{1} + \sin\theta_{31}\cos\theta_{32}e^{i\phi_{32}}\ket{2}\\
                     & + \sin\theta_{31}\sin\theta_{32}\ket{3},\\
              \vdots & \\
      \ket{\psi_n} = & \cos\theta_{n,1} \ket{1} + \sin\theta_{n,1}\cos\theta_{n,2} e^{i\phi_{n,2}} \ket{2} + \ldots \\
             & + \sin\theta_{n,1}\sin\theta_{n,2}\ldots\cos\theta_{n,n-1}e^{i\phi_{n,n-1}}\ket{n-1}\\
             & + \sin\theta_{n,1}\sin\theta_{n,2}\ldots\sin\theta_{n,n-1}\ket{n}.
 \end{split}
\end{equation}
From $\ket{\psi_{n+1}}$ on, the states have the full number of non-fixed parameters as given in Eq.~(\ref{eq:staterepres}).
Thus the first $n$ states of the quorum have $0,1,3,5,\ldots,2n-3$ free parameters, while each of the remaining states has
$2n-2$.

\subsection{Determinant of quorum as quality measure}
We consider the matrix $\mathcal{Q}$, which is formed by writing the operator $\mathcal{Q}_i$,
or rather its components when denoted in a orthogonal basis for the space of traceless $n\times n$ matrices,
as $i$th
row of $\mathcal{Q}$.
Then, the value of $|\det\mathcal{Q}|$ serves as a quality measure for the quorum.
It is identical to the volume of the parallelepiped spanned by the vectors corresponding to the $\mathcal{Q}_i$
in the $(n^2{-}1)$-dimensional vector space.
Wootters and Fields \cite{WoottersFields89} have shown that this volume evaluates
how much knowledge about an unknown quantum state can be obtained with a finite number of measurements.
Note that there exists the alternative approach of using the condition number of the reconstruction matrix as a quality measure
for the measurement set \cite{FilippovManko2010,Bogdanov2010,Miranowiczetal2015,Bartkiewiczetal2016}.
The determinant might be computed by just applying a simple Gaussian diagonalization scheme.
This means, for each step $k>0$
\begin{eqnarray}
 \text{for $i>k$:} ~~~\mathcal{Q}_i^{(k)} & = & \mathcal{Q}_i^{(k-1)} - \frac{ B_{ki}^{(k)} }{ B_{kk}^{(k)} } \mathcal{Q}_k^{(k-1)},\\
                              B_{ij}^{(k)} & = & \langle \mathcal{Q}_i^{(k-1)}| \mathcal{Q}_j^{(k-1)} \rangle_M,
\end{eqnarray}
where $\langle A|B \rangle_M = \sum_{i=1}^{n^2-1} A_i^*B_i$ is the dot product for the reduced projection vectors $\mathcal Q_j$.
Then the determinant is given by
\begin{equation}
\label{eq:det}
 \det\mathcal{Q} = \prod_{k=1}^{n^2-1}\sqrt{B_{kk}^{(k)}}.
\end{equation}
This method has the advantage that it is actually not necessary to calculate the $\mathcal{Q}_i^{(k)}$, because for $k=1$,
we have
\begin{equation}
 B_{ij}^{(1)}  =  |\langle \mathcal{Q}_i| \mathcal{Q}_j \rangle_M |= \langle\psi_i|\psi_j\rangle|^2 - 1/n
\end{equation}
and then we find as all the $B_{ij}^{(k)}$ are real
\begin{equation}
 B_{ij}^{(k{+}1)} = B_{ij}^{(k)} - \frac{ B_{kj}^{(k)}B_{ik}^{(k)} }{ B_{kk}^{(k)} }.
\end{equation}
However, this can include division by very small numbers which is numerically problematic,
therefore we use in practice more stable standard methods for calculating the determinant
using existing linear algebra libraries.
In order to do so, we define a basis in the $(n^2{-}1)$-dimensional matrix space where we then calculate
the matrix $\mathcal{Q}$.

\subsection{Formulation of optimization problem}
The remaining problem, which will be tackled numerically in the following section,
can be formulated as follows.
The function $D=|\det\mathcal{Q}|$ should be minimized as a function of the $N_{\rm param}$ parameters
\begin{equation}
\begin{split}
 & \theta_{21}, \theta_{31},\ldots,\theta_{n^2{-1},1},\theta_{32},\ldots,\theta_{n^2{-}1,2},\ldots,\ldots,\theta_{n^2{-}1,n{-}1},\\
 &   \phi_{32},   \phi_{42},\ldots,  \phi_{n^2{-}1,2},  \phi_{43},\ldots,  \phi_{n^2{-}1,3},\ldots,\ldots\phi_{n^2{-}1,n}.
 \end{split}
\end{equation}
The values for the $\theta_{mj}$ can be restricted to the interval $[0,\pi/2]$
and the parameters $\phi_{mj}$ can be restricted to $[0,2\pi)$.
For computing $D(\theta_{21},\ldots,\phi_{n^2-1,n})$, numerical standard methods for computing a determinant are applied.

\section{Numerical optimization}
\label{numerics}

\subsection{Methods}
\label{sec:method}
Due to the lack of information about the function to be optimized, an exploratory analysis was performed.
The ruggedness of the function, as well as parameter interdependency, were presumed.
As part of the exploratory analysis different methods from the optimx and optim
packages from R \cite{optimx1,optimx2}, were used.
The methods include local as well as global optimization approaches.
The Nelder-Mead or downhill simplex method\cite{neldermead},
a variable metric method, BFGS, which is based on \cite{BFGS},
"CG", which implements a conjugate gradients method based on \cite{CG1,CG2}, 
newton-like method for unconstrained problems with at least first derivatives, nlm \cite{nlm},
spg \cite{spgnew}, a non-monotone spectral projected gradient method, which
is based on \cite{spg1,spg2},
a quasi-Newton type general purpose optimization algorithm, ucminf, \cite{ucminf},
the method "optim:sann" \cite{sann}, which is a variant of simulated annealing,
belonging to the class of stochastic global optimization methods.
Powell's methods newuoa  \cite{powell2006newuoa}
and uobyqa \cite{powell2002uobyqa}, from the package optimx, were also tested.
For all algorithms the default stopping criteria were used,  the iteration
maximum was set to 10000.
Relative convergence tolerance was used as a stopping criterion, namely the
algorithm stops if it is unable to reduce the value $\mathtt{val}$ by a
factor of $\mathtt{reltol} * (\operatorname{abs}(\mathtt{val}) + \mathtt{reltol})$
at a step, where the relative tolerance $\mathtt{reltol}$
used was $1.490116\times10^{-8}$.

It was determined that, without special tuning, Powell's derivative-free methods performed the best for this problem.
The chosen method NEWUOA (NEW Unconstrained Optimization Algorithm) \cite{powell2006newuoa} was then applied
to solve the optimization problem.
NEWUOA is a derivative-free algorithm, which is based on a trust region technique when searching for the
optimal solution.
At each iteration, the algorithm uses quadratic interpolation to compute the objective function and then
performs conjugate gradient minimization within a trust region.
It then updates either the current best point or the radius of the trust region, based on the a posteriori
interpolation error. 

For state spaces of dimensions $n=3$ to $n=8$, 15 random points were used as starting points and the NEWUOA search was run. 
For $n=3$, additionally the algorithm was run with 2000 random starting points in order to compare it with
the theoretical hypothetical bound.
For finding the top result, the best results were repeatedly used as starting points until convergence.

Among the quorums which were found by numerical optimization with random starting points, there were some which had the property
that some of the obtained parameters were nearly identical or close to zero.
In order to make use of this, we implemented a modified search
\footnote{This optimization is implemented in python using the method "Powell" from the optimization package from the scipy library \cite{Scipy}.
          This method is a modified version of \cite{Powell1964}.
          We set the tolerance, which is allowed within the convergence criterion, of the value of the function to $10^{-17}$ in order obtain results close to the optimum.
          The allowed tolerance for the input parameters was set to $10^{-3}$.}
setting the similar parameters to be exactly identical and the ones close to zero to be exactly zero simplifies the problem.
Then, rerunning the optimization
with the reduced number of free parameters
using 25 random starting points, allowed us to find improved results.

\subsection{Results and discussion}
The optimized values for $|\det{\mathcal{Q}}|$ are presented below in Table \ref{tbl:comparison}.
Expectedly, the numerical optimization performs better for a lower number of free parameters,
finding nearly the same value of $|\det{\mathcal{Q}}|$ at each run for $n=3,4,5$,
while there is a larger variance for higher dimensions.
Consequently, the chances get higher that the best result out of the 15 runs which were performed
is still significantly below the global maximum the higher the value of $n$.
In the following, we will analyze the structure of the obtained optimized quorum
within the space for the traceless part of the measurement operators for three and four dimensions.
I order to do so, we calculate pairwise the absolute squared scalar products of the quorum states, which are directly related to the
angles between the respective vectors in operator space.
We show that the optimal quorum which was found numerically is not unique for four dimensions.
For three dimensions we will present analytical expressions for the parameters of the state of one quorum which
we assume to be optimal and up to permutations unique as it could not be improved numerically.
The parameters of the best performing quorums for $n=4,5,6$ are given in the Appendix.

Furthermore, we will discuss the robustness of the results against deviations in the measurement setup four $n=4$.
In case in an experiment, the projections are not precisely on the desired states of the optimized quorum, but the deviation is known,
the robustness given here allows to estimate the loss of performance.

\subsubsection{Three dimension}
Following the startegy described at the end of Sec.~\ref{sec:method},
we found a quorum with absolute squares of the quantum states, $W^{3D}_{ij} = |\langle\psi_i|\psi_j\rangle|^2$,
being, up to permutations, close to the following rational values,
\begin{equation}
 W^{3D} = \left(\begin{array}{cccc}
                W_1^{3D} & W_2^{3D} & W_2^{3D} & W_2^{3D} \\
                W_2^{3D} & W_1^{3D} & W_2^{3D} & W_2^{3D}\\
                W_2^{3D} & W_2^{3D} & W_1^{3D} & W_2^{3D} \\
                W_2^{3D} & W_2^{3D} & W_2^{3D} & W_1^{3D}
               \end{array}\right)
\end{equation}
with
\begin{equation}
 W_1^{3D} = \left(\begin{array}{cc}
                   1 & 4/9\\
                   4/9& 1
                  \end{array}\right)
\text{ and }
W_2^{3D} = \left(\begin{array}{cc}
                  7/27 & 7/27\\
                  7/27 & 7/27
                 \end{array}\right).
\end{equation}
Indeed we were able to identify the following parameters which provide exactly this quorum,
\begin{equation}
 \begin{split}
& \theta_{11} = 0, \\
& \theta_{21}=\arccos(-2/3),\\
& \theta_{31}=\theta_{41}=\theta_{51}=\theta_{61}=\theta_{71}=\theta_{81}=\arccos\sqrt{7/27},\\
& \theta_{12}=\theta_{32}=\theta_{72}=0,\\
& \theta_{22}=-\pi/3,\\
& \theta_{42}=\theta_{52}=\theta_{62}=\theta_{82}=\arcsin(3/4),\\
& \phi_{12}=\phi_{72}=0,\\
& \phi_{22}=-\phi_{42}=\phi_{52}=\phi_{62}=-\arccos(-1/(2\sqrt{7})),\\
& \phi_{32}=\arccos(-13/14),\\
& \phi_{82}=\arccos(10/(7\sqrt{7})),\\
& \phi_{13}=\phi_{33}=\phi_{43}=\phi_{73}=0,\\
&\phi_{23}=-\arccos\sqrt{3/7}+\arccos(-13/14),\\
& \phi_{53}=-\arccos(-1/(2\sqrt{7})),\\
& \phi_{63}=\pi/3,\\
& \phi_{83}=-\arccos(-1397/1778),
\end{split}
\end{equation}
providing $|\det\mathcal{Q}|=.158766448204$.

This value of $|\det\mathcal{Q}|$ is larger than of all numerically optimized with 2000 random starting points.
Therefore, we assume that it is indeed the optimal choice.
A proof of this assumption, however, is not provided here.
Note that other quorums which were found numerically contain some significantly different values for
$|\langle\psi_i|\psi_j\rangle|^2$ while $|\det\mathcal{Q}|$ is only a little bit lower.
For example, the quorum given by the parameters
\begin{equation}
\begin{split}
 (\theta_{21},\ldots,\theta_{81})  =& (1.035269, 2.033255, .887359, 1.037837,\\
                                    &.921986, 1.035269, 1.037836),\\
 (\theta_{32},\ldots,\theta_{82})  =&(.4920704, 1.283451, .8515625, .414486,\\
                                    & 2.295302, .8515648),\\
 (\phi_{32},\ldots,\phi_{82})=&( 6.8892, 6.052385, 4.53376, 2.208782,\\
                                     &4.148872, 1.749426),\\
 (\phi_{43},\ldots,\phi_{83}) =& (4.272501, 5.449324, 4.991345, .2010878,\\
                                     &2.664994)
\end{split}
\end{equation}
yields
\begin{equation}
\begin{split}
 &W^{3D}= \\
 &\left(\begin{array}{cccccccc}
   1.     & .2604 & .199 &  .3987 & .2581 & .3651 & .2604 & .2582\\
  .2604 & 1.      & .2581&  .2581  &.2604  &.2581  &.4445 & .2604\\
  .199 &  .2581 & 1.    &  .2604 & .3987  &.2604  &.2581 & .3651\\
  .3987 & .2581  &.2604  &1.      &.3651 & .2604 & .2581 & .199 \\
  .2581 & .2604  &.3987  &.3651 & 1.     & .199  & .2604 & .2581\\
  .3651 & .2581  &.2604  &.2604 & .199  & 1.     & .2581  &.3987\\
  .2604 & .4445  &.2581  &.2581 & .2604  &.2581 & 1.      &.2604\\
  .2582 & .2604  &.3651  &.199  & .2581  &.3987 & .2604  &1.    
               \end{array}\right)
\end{split}
\end{equation}
containing several values, namely $.199$, $.3987$, and $.3651$,
which differ severely from $4/9=.4444$ or $7/27=.2593$.
However, $|\det\mathcal{Q}|=.158766446951$ differs only in the in the order $10^{-9}$
from the value found for the assumed-to-be-optimal quorum given above.

\subsubsection{Four dimensions}
The optimized quorums, which were found for 15 different random starting points,
have nearly identical values for $|\det\mathcal{Q}|\approx.07843$.
Interestingly, for all of them the states of the quorum, $\ket{\psi_1},\ldots,\ket{\psi_{n2{-}1}}$, can be ordered in a way that the
symmetric matrix $W^{4D}$, defined as $W_{ij}^{4D} = |\langle\psi_i|\psi_j\rangle|^2$, takes the form
\begin{equation}
 W^{4D} = \left( \begin{array}{ccc} W^{4D}_1 & W^{4D}_2 & (W^{4D}_2)^T \\
                                    (W^{4D}_2)^T & W^{4D}_1 & W^{4D}_2\\
                                    W^{4D}_2 & (W^{4D}_2)^T & W^{4D}_1
                 \end{array}\right)
\end{equation}
with
\begin{equation}
 W^{4D}_1 = \left(\begin{array}{ccccc}
              1 & .2966 & .2966 & .1795 & .1795\\
              .2966 & 1 & .1795 & .2381 & .1921\\
              .2966 & .1795 & 1 & .1921 & .2381\\
              .1795 & .2381 & .1921 & 1 & .2966\\
              .1795 & .1921 & .2381 & .2966 & 1
             \end{array}\right)
\end{equation}
and
\begin{equation}
 W^{4D}_2 = \left(\begin{array}{ccccc}
              .1636 & .1921 & .1921 & .1921 & .1921 \\ 
              .2381 & .1677 & .2624 & .2624 & .1978 \\ 
              .2381 & .2381 & .2624 & .1677 & .1978 \\
              .2381 & .1677 & .1755 & .2624 & .1677 \\ 
              .2381 & .1755 & .1677 & .1677 & .2624
             \end{array}\right)
\end{equation}
where a deviation of the value is maximally $10^{-4}$.
This means that the operators of the quorum are alway arranged in the same way and the respective vectors show some structure.
Namely, there are three groups of states with the same relation towards each other within the group.
However, the quorum is not unique even when permutations are disregarded as the states are not equivalent, i.e. there are different ways to
arrange the construction given by $W^{4D}$ in the four-dimensional Hilbert space of the quantum states.

If the structure or some of its properties were known beforehand, one could formulate the optimization problem with less free parameters,
as some of them would be known to be identical.

\subsubsection{Robustness}
\begin{figure}
 \includegraphics{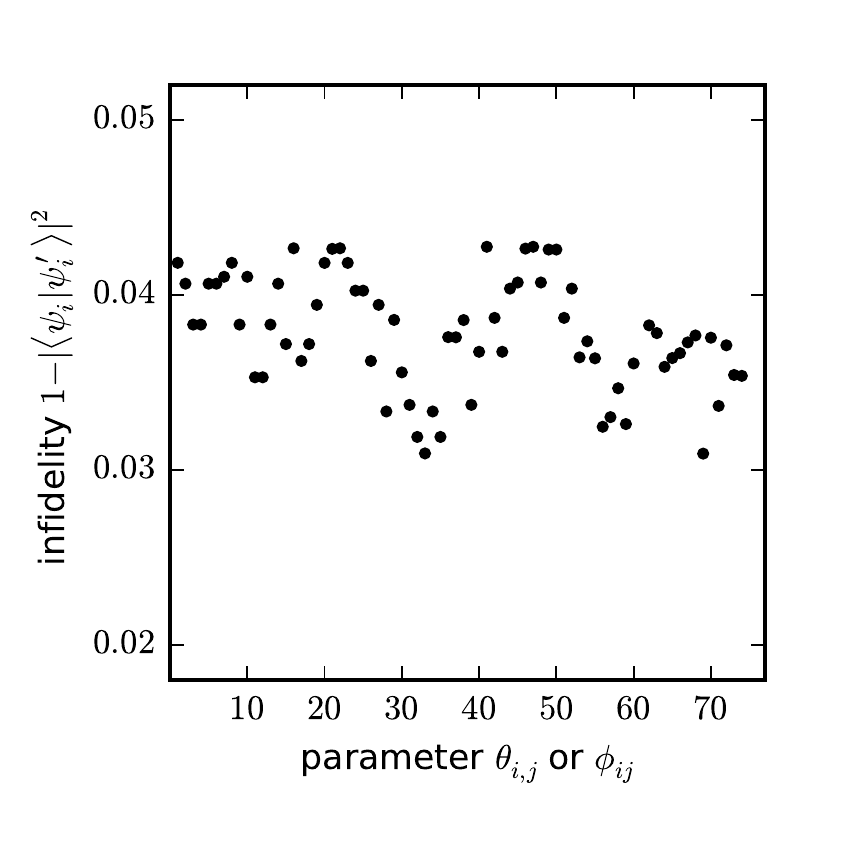}
 \caption{State infidelity $1 - |\langle\psi_i|\psi_i'\rangle|^2$ which corresponds to a shift of the parameter,
          $\theta_{ij}+\Delta\theta_{ij}$ or $\phi_{ij}+\Delta\phi_{ij}$ that reduces $|\det\mathcal{Q}'|$ by $5\%$
          compared to the desired $|\det\mathcal{Q}|$.
          The parameters are ordered as follows,
          $\theta_{21},\theta_{31},\ldots,\theta_{15,1},\theta_{32},\theta_{42},\ldots,\theta_{15,3},\phi_{32},\ldots,\phi_{15,4}$.
          The visible fluctuations are due to the specifications of the specific quorum chosen.
          It is given in the Appendix.
          Note that changes in the parameter $\phi_{12,3}$ do not yield a change in the determinant of $\mathcal{Q}'$ larger than 0.045
          as $|\langle\psi_{12}|3\rangle|$ happens to be rather small.
          Therefore, this parameter is left out in the figure above.}
 \label{fig:robustness4d}
\end{figure}

We consider the robustness by calculating $|\det\mathcal{Q}'|$ for $\mathcal{Q}'$ being the quorum one obtains by a shift
$\theta_{ij}\to\theta_{ij}+\Delta\theta_{ij}$ or $\phi_{ij}\to\phi_{ij}+\Delta\phi_{ij}$ keeping the other parameters
constant.
We calculate the positive and the negative values for $\Delta\theta_{ij}$ and $\Delta\phi_{ij}$ which correspond to $|\det\mathcal{Q}'|$
being $5\%$ reduced compared to $|\det\mathcal{Q}|$ and present the averaged respective state infidelity $1 - |\langle\psi_i|\psi_i'\rangle|^2$
where $\ket{\psi_i'}$ is the state for the shifted parameter, $\theta_{ij}+\Delta\theta_{ij}$ or $\phi_{ij}+\Delta\phi_{ij}$,
in Fig.~\ref{fig:robustness4d}.
We show here only the robustness for a quorum in four dimension, which we consider to be the most relevant case as it corresponds to two qubits.
Note that in an experiments the shifts $\Delta\theta_{ij}$ and $\Delta\phi_{ij}$ have to be known in order to apply the results shown here.
Uncertainties, i.e., noise, has to be taken into account differently.
However, Fig.~\ref{fig:robustness4d} shows that even if projections on a quantum state deviates from the desired quorum state by a state infidelity of around $4\%$, the
"volume" of the set of measurement in $(n^2{-}1)$-dimensional space of the traceless parts of its projections is only slightly ($5\%$) reduced compared to a gain of more than
a factor of two compared to the quorum from MUBs, see Sec.~\ref{compare}.
Thus, small imperfections of its realization have less influence than the overall choice of the quorum itself.

\section{Comparison to a set from MUBs}
\label{compare}
If a set of $n+1$ MUBs exist, which is certainly the case for $n$ being an integer power of a prime number \cite{WoottersFields89},
a quorum can be formed by choosing $n-1$ states from each MUB, resulting in $n^2-1$ states in total.
The states from different MUBs are unbiased, i.e., the corresponding row vectors in the matrix $\mathcal{Q}^{\rm (MUB)}$ are diagonal.
Thus, we can write $\mathcal{Q}^{\rm (MUB)}$ as a block diagonal matrix, where we just need to diagonalize the blocks,
which are $(n-1)\times (n-1)$ matrices.
This is a rather simple task, because we know already that the corresponding quantum states are diagonal.
Therefore, $\langle \mathcal{Q}^{\rm(MUB)}_i | \mathcal{Q}^{\rm(MUB)}_j \rangle_M = 1/n$
if $\mathcal{Q}^{\rm(MUB)}_i$ and $\mathcal{Q}^{\rm(MUB)}_j$ are from the same basis and $i\neq j$.
Using the same diagonalization scheme as above for the block and also adapting the notation,
we obtain by straightforward calculation
\begin{equation}
 B_{ik}^{(k)} = \frac{n-k}{(n-k+2)(n-k+1)}~~\text{ if }i>k
\end{equation}
and
\begin{equation}
 B_{kk}^{(k)} = \frac{n-k}{n-k+1}
\end{equation}
Then the absolute value of the determinant can be expressed as
\begin{equation}
 |\det\mathcal{Q}^{\rm (MUB)}| = \left(\! \frac{(n{-}1)(n{-}2)\ldots1}{n(n{-}1)\ldots2} \!\right)^{\frac{n+1}{2}} = \frac{1}{n^{(n{+}1)/2}}.
\end{equation}
\begin{table}
 \begin{tabular}{|l| l| l| l|}
 \hline
  $n$ & $|\det\mathcal{Q}^{\rm MUB}|$       &  $\max|\det\mathcal{Q}|$ &  bound    \\
  \hline
  3   & $\frac{1}{9}$                       &  $.1588$                & $.1975$ \\
  \hline
  4   & $\frac{1}{32}$                      &  $.07843$               & $.1156$ \\
  \hline
  5   & $8\times 10^{-3}$                   &  $.04076$               & $.06872$\\
  \hline
  6   & {\color{gray}$1.8900\times10^{-3}$} &  $.02140$               & $.04115$\\
  \hline
  7   & $4.1650\times10^{-4}$               &  $.006313$              & $.02473$\\
  \hline
  8   & $8.6317\times10^{-5}$               &  $.001803$              & $.01490$\\
  \hline
 \end{tabular}
 \caption{Comparison of the results for dimension $n$ for the MUB quorum and for the numerically optimized quorum.
          Note that for $n=6$ a set of MUBs is not known.
          The numerical results (third column) clearly outperform the MUBs.
          We also provide an upper bound, which follows from the length of the row vectors in $\mathcal{Q}$.
          The results are visualized in Fig.~\ref{fig:num_MUB}}
  \label{tbl:comparison}
\end{table}
\begin{figure}[ht]
 \includegraphics{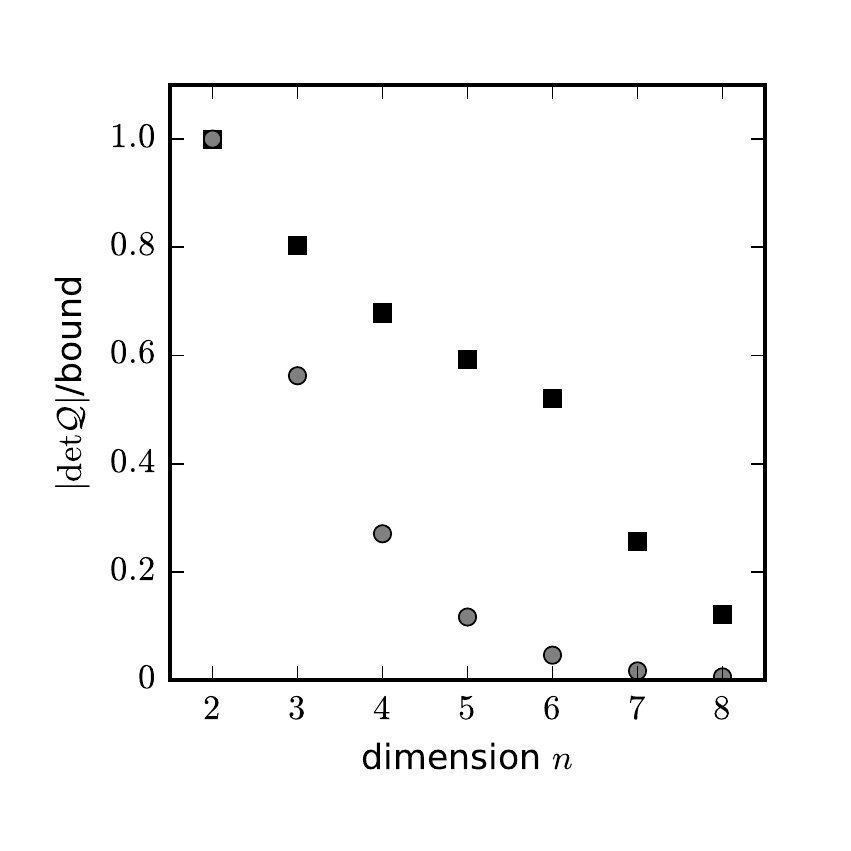}
 \caption{Values of $|\det\mathcal{Q}|/$bound for the numerically optimized quorums (black squares) and the quorums constructed from MUBs (gray circles).}
 \label{fig:num_MUB}
\end{figure}
In Table \ref{tbl:comparison} and Fig.~\ref{fig:num_MUB}, we compare the results obtained in the previous section to the result one would get
for a MUB quorum and to the upper bound $((n-1)/n)^{(n^2-1)/2}$.
The bound just follows from the length of the row vectors in $\mathcal{Q}$, $\sqrt{(n-1)/n}$.
Note that this bound cannot be reached, which was shown explicitly for $n=4$ in \cite{RohlingBurkard2013}.
We assume that this is the case for all dimensions $n>2$.
The improvement of the numerical optimization compared to the quorums from MUBs becomes more significant with increasing dimensionality.
Note, however, that for a large Hilbert space it is rather unpractical to perform state tomography by projections on individual states.

\section{Conclusions and Outlook}
\label{conclusionsandoutlook}

In this article, we have presented numerically optimized quorums for state tomography
based on measurement operators which are projectors on individual quantum states
in contrast to the more common non-degenerate measurements.
The results are clearly improved compared to a quorums constructed from states taken from sets of MUBs.
The best quorums for three and four dimensions show an interesting arrangement of the states.
Analyzing the structure of optimized quorums in higher dimensions and investigation how optimal quorums can be constructed
rather than numerically found, is beyond the scope of this article and should be the objectives of further studies.

Our approach, to apply numerical optimization in order to determine a good choice for a quantum state tomography scheme,
can be extended to models which include noise, i.e.,  the measurements are not perfect and the performance of each measurement migth be different.
Furthermore, one can include the quantum gates, i.e., the unitary transformations, which needs to be applied prior to the actual measurement,
and the imperfections to those gates in the optimization.
This means the approach would not only provide the quantum states which form an optimal quorum
but also the operations which are necessary to perform these measurements and minimize the expected uncertainty
of the tomography scheme.
Another direction for future research are degenerate measurements which are different from the projections on individual quantum states
considered here, e.g., the case $n=4$ with projections on two-dimensional subspaces which refers to two qubits where one of them is measured.
In combination with previously applied quantum gates this can also provide full state tomography of the two-qubit system.

Generally speaking, our optimization approach can be extended in order to provide customized tomography schemes
for experimentally realized quantum systems and measurement setups.

\section*{Acknowledgements}
We thank Guido Burkard for helpful discussions.
This work was partly supported by
the Research Council of Norway through its Centers of Excellence funding scheme, project number 262633, "QuSpin".

\onecolumngrid
\appendix*
\section{Parameters of the best performing quorum from the numerical optimization}

Here, we give the numerically determined parameters of the optimization problem belonging to the best results obtained for $|\det\mathcal{Q}|$
for the dimensions four, five, and six.
For seven and eight dimensions we spare the sets of parameters due to their length.
Note that those parameters which we have chosen to set to zero are not given again.
The parameters are ordered starting in a way that we first present $\theta_{i1} (i=2,\ldots,n^2-1)$, then the other
$\theta_{ij} (j=2,\ldots n-1)$ and then the $\phi_{ij}$ until $\phi_{n^2{-}1,n}$.
Note that the restriction of $\theta_{ij}$ to $[0,\pi/2)$ and of $\phi_{ij}$ to $[0,2\pi)$
was applied to the starting points but not enforced on the optimization.
Thus, some of the values given here, are outside this interval.

{\it four dimensions:} $|\det\mathcal{Q}| = .0784336423365$
\begin{equation}
 \begin{split}
  &(\theta_{21},\theta_{31},\ldots,\theta_{15,1},\theta_{32},\theta_{42},\ldots,
  \theta_{15,3},\phi_{32},\ldots,\phi_{15,4})\\
 = & (-1.987238, 1.11715, 2.0805, 2.080499, 1.117151,   1.11715, 2.008353, 1.987238, 2.080499, 2.008353,   2.1467, 5.288292,\\
   &   2.080499, 2.024443, 2.495647, 1.119342, .9914846, 2.495645, 1.10621, .9093897,  1.862942, 1.11934, 2.232204, 2.247469,\\
   &   2.247471,  2.15011, 1.106211, 2.915391,  2.898041, 1.031702, .825679, 1.133683, .6625359, 2.239346, .5620509, 1.431614,\\
   &  2.519742,   1.004276, 1.933486, 6.554038,   7.994821, 1.967422, 6.012345, 7.279308, .98654, 6.283182, 4.571561, 2.155061,\\
   &  4.287843, 1.995348,   7.457368, 2.145481, 5.064952, 3.087178, 3.03431,  4.531637, 6.351414, 4.985414, 4.524217, 4.090605, \\
   &   6.983559, 3.394178, 2.692911, .7303369,  .651116, 2.154539, 7.25511, 4.092227, 2.967463, 6.387222, 1.057313, 1.975125,\\
   &  8.553043, 2.356182, 6.866927)
 \end{split}
\end{equation}

{\it five dimensions:} $|\det\mathcal{Q}| = .0407645110122$
\begin{equation}
\begin{split}
 &(\theta_{21},\theta_{31},\ldots,\theta_{24,1},\theta_{32},\ldots,\theta_{24,4},\phi_{32},\ldots,\phi_{24,5})\\
 = & ( 2.025332, 1.063227, 1.187547, 1.954049, 1.184432, {-}1.149042, 2.014489, 1.964083, 2.02577, 1.185496,   1.121085, 1.086572,\\
   & 1.161708, 1.139781, 2.055027,  1.169922, 1.178831, 1.160673, 1.115826, 1.168168,  1.187548, 2.024097, 2.065788, 2.101547,\\
   & .9118843,   1.234142, .8088603, 1.219918, 2.324767, 1.131091,   1.968451, 1.121148, 1.05738, 2.086829, .9157541,   1.915946, \\
   & 1.84196, .9293043, 1.202568, 1.083903,  1.797014, 1.259305, 1.019004, 1.198495, 2.017589, 1.951278, .9733781, 1.262492,\\
   & .923137, 1.845973, .9209366, 1.779616, 1.015688, .6300233, 1.366076, .915297, 1.302166, 1.409244, .7888966, .5972418,\\
   & .6029987, .8702579, 2.064999, .9712894, 1.105447, 2.211262, 2.012101, 1.296047, .5692779, 1.367542, -.5053642, 1.052135,\\
   & .3175097, 1.141402, .3709571, .3776104, .7486942, .7060893, 2.734705, .8864123, 2.846409, -1.122271, .7982444, -1.088667,\\
   & 2.669574, 2.04523, 8.225065, 6.850108, 6.277125, 5.770382, 5.008554, 6.695896, 4.893754, 3.96695, 1.531828, 4.747936,\\
   & 1.806821, 5.248359, 1.065362, 4.062266, 5.003858, 3.000425, 1.328327, 6.302228, 3.702629, .9964051, 6.258343, 8.023806,\\
   & -.9168062, .9622705, 2.20119, 2.195767, -.4585056, -.3640269, .3979758, 1.388356, 2.117641, 2.005458, 4.532796, 4.856528,\\
   & 2.372003, 4.202955, 1.991998, -1.257368, 3.796135, 3.683472, .2753047, 3.692093, .6109305, 7.674062, 9.065004, 2.675162,\\
   & -1.163622, 5.801429, 1.455933, 4.373961, 4.228544, 7.097044, 3.308416, 4.137023, 4.796179, 2.713446, 7.061001, 5.772213,\\
   & .5179369, 5.173816, 7.975781, 2.503486, 6.562516, 5.218709, 5.541723, 4.674475, .8936102, 5.298166, .7285783, .6042217,\\
   & 2.206226, 2.48539, 6.088396, 2.542696, 4.578949, 3.910954, 6.342585, 3.870752, 4.560328, 1.161769, 6.149991, 3.581524)
 \end{split}
\end{equation}

\newpage
{\it six dimensions:} $|\det\mathcal{Q}|= 0.02180422$

\begin{equation}
\begin{split}
  & (\theta_{21},\theta_{31},\ldots,\theta_{35,1},\theta_{32},\ldots,\theta_{35,5},\phi_{32},\ldots,\phi_{35,6})\\
= & (1.200726, 2.007267, 1.136718, 1.212113, 1.166749, 1.955048, 1.938887, -8.285738, 2.021419, 1.956438, 1.168484, 1.930938,\\
  & 2.025287, -4.354848, 1.941466, 2.007702, 1.959066, 1.967186, 1.943287, 1.95033, 1.149914, 1.183036, 1.216544, 1.19203,\\
  & 1.204242, 1.959146, 1.174527, 1.194205, 1.141995, 1.952349, 1.927501, 1.211222, 1.208198, 1.12139, 1.064684,  2.086422\\
  & 1.9439, 2.201574, 1.863483, 1.131354, 1.148588, .8711998, 1.937272, 1.925475, 1.206975, 1.897486, 2.127146, 1.868534,\\
  & 1.837702, 2.207828, 1.03747, 1.96676, 2.203703, 1.964852, 1.127361, .9741118, 2.076148, 2.078879, 7.602875, 2.016043, \\
  & 1.823238, 1.866946, 1.247623, 1.258312, 2.068498, 1.207911, 2.069339, 1.771407, .9804647, 1.157219,  1.103222, 1.991693,\\
  & .9509069, .6588244, 2.063136, -1.146611, 2.257391, 1.957527, 1.191194, 1.204787, 1.197062, 1.63133, 1.992722, .7632899,\\
  & 1.231786, 1.343243, 1.716956, -0.8128132, 1.502357, 1.182634, 1.2321, 2.205136, .8250355, 2.176076, .9415523, 2.304861, \\
  & .8823443, 1.887888, 1.367516, 1.030026, 1.652564, 2.012481, 2.232824, 2.224815,1.251554, 2.428719, 2.121333, 2.157949,\\
  & 1.208416, 1.183949, 2.338161, 1.217264, 1.648025, 1.038377, .9404738, 2.555162, 2.605002, .9860226, .6911011, 1.718876,\\
  & 1.382294, 2.057402, 2.448013, 1.05691, 1.170197, .9376441, 1.063722, 2.420602, .900834, .9794334, 1.520725, 2.072939,\\
  & -0.116024, 2.516558, 4.064503, 2.088605, 2.543337, 2.422057, .9362635, 1.141614, 1.860171, .8277882, 2.369419, .8202935,\\
  & 1.656815, 3.433743, .5829571, 1.119176, 2.65132, -0.7378547, .4070498, .7148053, 2.42493, 1.423588, .3030078, 1.844847,\\
  & -0.3382007, 3.890063, 2.028514, .4060558, 4.904644, 4.90699, 1.696743, 6.200229, 5.828303, 7.514629, 4.948504, -0.119077,\\
  & 4.79245, 2.538788, 1.588528, .1839321, 3.898752, 6.239238, -0.3239425, 3.202382, 7.133572, 7.715874, 3.939151, 4.72454,\\
  & 1.709308, 8.44531, 4.641578, 4.920483, 3.364537, 1.680187, 3.172782, 4.011768, 2.640836, 3.799028, .9947096, 7.476654,\\
  & -0.9945264, 4.38951, 3.769075, 2.161899, 3.928993, 4.267702, 1.647526, 4.847141, 4.082247, 3.027623, 2.474754, 3.93895, \\
  & 2.47395, -0.05955911, 4.831441, -1.871281, 5.520931, 2.018618, 1.692524, 8.043, 6.460713, 5.610878, 2.603571, 4.934541,\\
  & 5.695705, 4.833942, -0.1214876, 3.862147, 1.26356, .5518539, .4031677, -1.23673, -1.565766, 1.632747, 6.026669, 4.477365,\\
  & 3.363691, 2.212875, 8.209061, 5.764668, 6.732687, 2.478291, 1.589077, 4.579485, 1.926466, 2.411238, 1.443562, 3.50158,\\
  & 1.333135 3.995756, 6.938622, 1.851612, 5.173889, 1.587738, 6.237971, 4.8003, 6.100522, 4.027243, 1.085882, 2.080531,\\
  & 4.132851, 2.844395, 1.822641, 4.423196, 4.160778, 2.260777, 5.634928, -0.09330294, 6.368033, 4.231467, 7.917326, 3.783284,\\
  & 4.940628, 4.534321, 9.482261, 2.510288, 3.609704, 4.197574, 4.131333, 5.933377, 3.558919, 5.804634, 8.318015, 3.207322,\\
  & 5.103573, 8.584231, 3.727574, -0.2379239, 7.491072, 6.295222, 6.716216, 8.036527, 3.512619, 7.393747, 5.192058, 6.201087,\\
  & 5.450548, 2.864755, 6.192834, 4.871876, 5.751969, 7.261904, 5.910425, 2.041802, 7.703962, 1.025542, 8.406518, 5.986424,\\
  & 2.717615, 4.374027, 3.573012, 3.021222, 4.182929, 5.589567, 4.574366, 5.864266, 5.576348, 3.837732, 3.23019, 4.32822,\\
  & 4.400782, 7.319795, .606244)
\end{split}
\end{equation}

\twocolumngrid

\end{document}